# Imaging reconfigurable molecular concentration on a graphene field-effect transistor

Franklin Liou[1,2,3†], Hsin-Zon Tsai[1,2†], Andrew S. Aikawa[1,2], Kyler C. Natividad[1], Eric Tang[1], Ethan Ha[1], Alexander Riss[4], Kenji Watanabe[5], Takashi Taniguchi[6], Johannes Lischner[7], Alex Zettl[1,2,3], and Michael F. Crommie[1,2,3]*

[1]Department of Physics, University of California at Berkeley, Berkeley, CA 94720, United States.

[2]Materials Sciences Division, Lawrence Berkeley National Laboratory, Berkeley, CA 94720, United States.

[3]Kavli Energy NanoSciences Institute at the University of California at Berkeley, Berkeley, CA 94720, USA.

[4]Physics Department E20, Technical University of Munich, James-Franck-Straße 1, D-85748 Garching, Germany

[5]Research Center for Functional Materials, National Institute for Materials Science, 1-1 Namiki, Tsukuba 305-0044, Japan

[6]International Center for Materials Nanoarchitectonics, National Institute for Materials Science, 1-1 Namiki, Tsukuba 305-0044, Japan

[7]Department of Materials, Imperial College London, Prince Consort Rd, London SW7 2BB, UK.

[†]F.L. and H.-Z. T. contributed equally to this paper

*Corresponding author: crommie@berkeley.edu.

**Abstract:**

The spatial arrangement of adsorbates deposited onto a clean surface in vacuum typically cannot be reversibly tuned. Here we use scanning tunneling microscopy to demonstrate that molecules deposited onto graphene field-effect transistors exhibit reversible, electrically-tunable surface concentration. Continuous gate-tunable control over the surface concentration of charged $F_4TCNQ$ molecules was achieved on a graphene FET at $T = 4.5K$. This capability enables precisely controlled impurity doping of graphene devices and also provides a new method for determining molecular energy level alignment based on the gate-dependence of molecular concentration. The gate-tunable molecular concentration can be explained by a dynamical molecular rearrangement process that reduces total electronic energy by maintaining Fermi level pinning in the device substrate. Molecular surface concentration in this case is fully determined by the device back-gate voltage, its geometric capacitance, and the energy difference between the graphene Dirac point and the molecular LUMO level.

**Introduction:**

One important way in which adsorbates modify surfaces is by inducing localized electronic defect states that trap electrons and cause Fermi level pinning.[1–4] Typically, Fermi level pinning is considered a "one-way" process where defects may change the location of electrons (i.e., by trapping them) but electrons do not change the location of defects. Here we reverse this idea by implementing a condensed matter system where defect concentration can be continuously and reversibly tuned by adding or removing electrons from the system. We observe this unique behavior for $F_4TCNQ$ molecules adsorbed onto the surface of a graphene field-effect transistor (FET). When a voltage is applied to the back-gate of such a device under proper conditions then the resulting electric field is not screened by the graphene Dirac band electrons, but is rather unexpectedly screened by ionized molecules that mechanically rearrange themselves on the device surface. The devices in which this occurs are not electrochemical cells[5–7] attached to some external reservoir of material. Instead they are composed of ultraclean monolayers of graphene on hBN that are dosed with a submonolayer molecular coverage and then held at cryogenic temperature in ultrahigh vacuum (UHV). The adsorbate rearrangement process observed on our devices arises from a physical mechanism whereby adsorbate-induced Fermi level pinning helps to minimize graphene Dirac band electronic energy. The energetics of Fermi level pinning in this 2D system is so strongly tied to the adsorbate geometry that it allows reversible, mechanical alteration of the surface defect concentration by adding or removing electrons using the FET back-gate.

This behavior arises due to the proximity in energy of the $F_4TCNQ$ lowest unoccupied molecular orbital (LUMO) to the graphene Dirac point.[8–12] The resulting high electron affinity of $F_4TCNQ$ on graphene has been exploited previously to p-dope graphene.[13,14] STM measurements have also shown that charge flows easily into and out of $F_4TCNQ$ molecules on graphene.[8,10,15,16] No previous studies, however, have demonstrated reversible control over the geometric arrangement of $F_4TCNQ$ adsorbates on graphene.

The strategy of this paper is to first explain the experimental procedure by which we control the surface concentration of $F_4TCNQ$ molecules on a graphene FET via electrical signals sent to the device. This technique has enabled us to discover that the $F_4TCNQ$ surface concentration on graphene FETs varies linearly with applied gate voltage. We have characterized this unexpected behavior by performing STM spectroscopy on molecule-decorated graphene devices for different molecular coverages. These measurements allow us to establish a connection between gate-dependent molecular surface concentration and Fermi level pinning in graphene FETs. We have distilled these observations into a simple theoretical model that connects the energetics of graphene electrons to the geometric arrangement of surface adsorbates, and which allows us to predict molecular surface concentration for any given back-gate voltage. The connection established here between molecular electronic structure and molecular surface concentration provides a new technique for quantitatively determining molecular energy levels by simply counting the number of molecules on a clean graphene surface.

**Results**

Our technique for reversibly changing molecular concentration on graphene devices starts with the deposition of a submonolayer coverage of $F_4TCNQ$ molecules onto a clean graphene/hBN FET held at room temperature in UHV (the transconductance of the device and an SEM image is shown in Fig. S1). We then cool the molecule-decorated device down to 4.5 K without breaking vacuum, at which point the molecules can be stably imaged by our STM. In order to set the molecular surface concentration to a desired value, a set-voltage ($V_{G\text{-set}}$) is applied to the device back-gate while a source-drain current ($I_{SD}$) is *simultaneously* passed through the graphene as sketched in Fig. 1 ($I_{SD}$ provides thermal energy to facilitate molecular diffusion). The molecules freeze in place as soon as $I_{SD}$ is set to zero, resulting in a well-defined and reversible surface concentration that is determined by $V_{G\text{-set}}$. After the molecules are frozen in place, the gate voltage is set to zero for STM imaging. The resulting control over molecular surface concentration can be seen in the differently prepared molecular concentrations shown in Figs. 2a-f (all measured at the same spot on the surface). At a high set-voltage of $V_{G\text{-set}} = 60$ V the resulting molecular surface concentration is correspondingly high ($n_M = 6 \times 10^{12}$ cm$^{-2}$) and the roughly evenly-spaced molecular distribution suggests that the molecules are charged during the surface concentration-setting process and repel each other (Fig. 2a). As $V_{G\text{-set}}$ is reduced, the resulting molecular concentration correspondingly reduces (Figs. 2b-f). A plot of molecule surface concentration vs. $V_{G\text{-set}}$ (Fig. 2g) shows that the molecular concentration is almost perfectly linear over the gate-voltage range -10 V < $V_{G\text{-set}}$ < 60 V and remains nearly zero for $V_{G\text{-set}}$ < -10 V. The gate-tunable molecular concentration is observed to be non-hysteretic since the forward sweep and backward sweep data lie almost perfectly on top of each other. The molecular concentration within our STM field of view is thus precisely and reversibly controlled by tuning the graphene FET electrical device parameters.

The molecular concentration across the entire device surface, however, cannot be completely uniform since the molecules have to *go somewhere* when their concentration is reduced and *come back from somewhere* when their concentration is increased, and this "somewhere" cannot have the same concentration as the STM field of view seen in Fig. 2.

However, the process described above for setting the molecular surface concentration is surprisingly robust and can be observed over a majority of the surface area that we have scanned (~80 %) on multiple devices. A fraction of the device surface is defective/contaminated (~20 %) (and therefore unsuitable for these measurements) since the devices are fabricated under ambient conditions before being placed in our UHV chamber for measurement. We hypothesize that defective regions of the surface act as *local reservoirs* that store $F_4TCNQ$ molecules in a compact, condensed phase. $F_4TCNQ$ molecules must flow in and out of these condensation regions (due to electrostatic potential inhomogeneities) to enable the precise concentration control that we observe over a large majority of the surface. The fact that we observed no decrease in molecule concentrations over multiple cycles of density changing operations (Fig. S2) suggests the molecular storage and release process is fully reversible. To explain why molecules can be aggregated in the condensed phase, we hypothesize that a highly negative gate voltage draws out electrons from the system and thus tends to remove charges from molecules, therefore without repulsive Coulomb interactions between molecules, attractive intermolecular forces (such as van der Waals) causes the condensed phase to become energetically favored. Due to the weak molecule-substrate interaction, the molecules diffuse easily to the defective regions where they can be captured and removed from the non-defective regions. A definitive test of this "local reservoir hypothesis" is difficult since mapping the entire surface of a graphene FET at the atomic scale is currently not possible (highly defective regions cannot be imaged), but we note that a condensed phase of $F_4TCNQ$ molecules has been observed on $F_4TCNQ$-decorated graphene FETs in the past.[17]

In order to understand the microscopic mechanism that underlies the process of tuning molecular surface concentration we must understand how charge readjusts itself in a graphene device decorated by mobile, charge-tunable molecules as the gate voltage is changed. Here the molecule-decorated graphene forms one side of a capacitor while the back-gate electrode forms the other, and so the total charge that flows to the molecule/graphene system can be determined straightforwardly from the device capacitance and applied gate-voltage ($Q_T = -V_{\text{G-set}} C_T$). Electrons transferred to the graphene side of the capacitor can either occupy graphene Dirac fermion band states or, alternatively, the LUMO states of the adsorbed molecules. To clarify how charge is distributed between these two choices we performed STM spectroscopy directly on adsorbed $F_4TCNQ$ molecules as well as on bare graphene patches between the molecules for different applied gate-voltages and different molecular concentrations. This enabled us to track the energetic alignment of the $F_4TCNQ$ LUMO energy ($E_L$) and the graphene Dirac point energy ($E_D$) with respect to the graphene Fermi level ($E_F$), thus providing snapshots of how charge in the device is distributed under different conditions. Our observations in this regard provide the basis for a simple model that quantitatively predicts gate-tunable molecular surface concentration.

We first describe the negative gate-voltage regime, $V_G < -10$ V, where the electronic configuration of the molecule/graphene system is qualitatively represented by the sketch in the inset to Fig. 3c. Here the graphene is p-doped since $E_F$ lies below $E_D$. The molecular LUMO level, $E_L$, lies below $E_D$ but remains unfilled by electrons since it is still higher than $E_F$. Experimental evidence for this type of electronic behavior is seen in the "on-molecule" and "off-molecule" $dI/dV_S$ spectra of Figs. 3a and 3b, where $V_S$ is the sample bias of the tunneling current (all $dI/dV_S$ spectra shown in Fig. 3 were acquired after "freezing" the molecules by setting $I_{SD}$ to zero -- spectroscopy was never acquired under "diffusive" conditions where $I_{SD} \neq 0$). Here molecular spectra were only measured in the range of positive $V_S$, since at negative $V_S$, tip electric fields can cause the molecules to charge and adsorb onto the tip. The top curve in Fig. 3a shows an on-molecule $dI/dV_S$ spectrum measured with the STM tip held above the center of an F$_4$TCNQ molecule for a gate-voltage of $V_G$ = -60 V and a fixed molecular surface concentration of $n_M$ = 0.8 x 10$^{12}$ cm$^{-2}$ (see inset). The F$_4$TCNQ LUMO level (arrow) sits at a sample bias of $V_S \approx$ 0.2 V above the Fermi energy (which is at $V_S$ = 0). The precise energy position of the LUMO is determined following the protocol of ref. [8] after accounting for known inelastic tunneling effects (see Fig. S3).[8,18,19] Specifically, a 35 meV phonon mode found by fitting high resolution STS measurements broadens the LUMO peak,[8] resulting in the actual LUMO energy being lower than the apparent maximum in the spectrum by 35 meV. Previously reported satellite peaks due to a 183 meV phonon mode is also visible at $V_S \approx$ 0.4 V.[8]

The top curve in Fig. 3b shows an off-molecule $dI/dV_S$ spectrum taken with the STM tip held over a bare patch of graphene ~100 Å away from any F$_4$TCNQ molecules for the same gate-voltage and molecular concentration used for the top curve of Fig. 3a. A depression in $dI/dV_S$ over the range -60 mV < $V_S$ < 60 mV can be seen that is known to occur due to phonon inelastic tunneling effects,[20] while another depression is seen at $V_S \approx$ 0.35 V that marks the location of the graphene Dirac point ($E_D$).[20,21] The precise value of $E_D$ is found by fitting the $dI/dV_S$ spectra using the protocol outlined in the Supplementary Information. The Dirac point is seen to fall in energy as the gate voltage is increased, causing the graphene to transition from being hole-doped ($E_D$ > $E_F$) to being electron-doped ($E_D$ < $E_F$) at $V_G \approx$ 20 V for this molecular surface concentration ($n_M$ = 0.8 x 10$^{12}$ cm$^{-2}$). $E_D$ (from Fig. 3b) is observed to lie above $E_L$ (from Fig. 3a) by ~150 mV for negative gate voltages, and both quantities shift downward in energy together as the gate voltage is increased from $V_G$ = -60 V to -10 V (Fig. 3c). Such behavior is expected as $E_F$ "rises" with increased gate voltage in the band structure shown in the inset to Fig. 3c.

When $V_G$ is raised above -10 V the molecules become charged, causing them to enter a new physical regime. The transition toward this regime can be seen in the gate-dependence of $E_L$, which falls towards $E_F$ with increasing $V_G$ in Fig. 3a. Another signature that the molecules become charged (i.e., that the LUMO becomes filled) is that they become mechanically unstable during STS measurement when $V_G$ is raised above -10V. Negatively charged molecules are observed to escape from under the STM tip during spectroscopy measurements, making it impossible to obtain reproducible "on-molecule" spectroscopy for $V_G$ > -10 V. Interestingly, this critical gate-voltage value coincides with the $V_{G\text{-set}}$ value at which the molecular concentration begins to rise from near zero (for increasing $V_{G\text{-set}}$), showing that the appearance of molecules on the device surface is correlated with their charge state. Another indication of charged molecular behavior is the spatially uniform intermolecular separation, a signature of intermolecular repulsion.

In order to better understand the charged molecular regime we investigated how changing the molecular surface concentration affects the way electrons fill up graphene states during electrostatic gating. This was accomplished by inspecting the gate-dependence of $E_D$ for different fixed molecular concentrations. Such measurements reveal charge transfer to $F_4$TCNQ LUMO levels and how this leads to Fermi-level pinning. To see this we first set the molecular concentration to a desired value (as shown in Fig. 4a) and then measured the gate-voltage dependence of $E_D$ (Fig. 4b) using off-molecule dI/dV$_S$ spectra via the procedure described above for Fig. 3b. For a pristine graphene capacitor, $E_D$ is expected to move smoothly down in energy with increasing $V_G$ according to the well-known expression[20]:

$$E_D(V_G) = -sgn(V_G)\hbar v_F\sqrt{\pi C|V_G - V_0|} \; , \tag{1}$$

where $v_F = 1.1 \times 10^6$ m/s is the electron Fermi velocity in graphene, C is the unit area capacitance of the device and $V_0$ reflects background doping. By fitting Eq. 1 to the gate-dependent Dirac point energy of our device before depositing molecules we are able to extract the capacitance between the graphene and the back-gate electrode: C = (7.8 ± 0.2) × $10^{10} cm^{-2} V^{-1}$ (Fig. S5).

For nonzero molecular concentrations, however, the $E_D$ vs. $V_G$ curve deviates from Eq. 1 and forms a "pinning" plateau at ~140 meV above $E_F$, with the width of the plateau increasing with increased molecular surface surface concentration (Fig. 4b). The start of the plateau (for increasing $V_G$) coincides with the gate voltage value where the molecular LUMO begins to fall under $E_F$, thus allowing us to associate the plateau with the charging of $F_4TCNQ$ molecules. The value of $E_D$ at the plateau (~140 meV above $E_F$) is very close to the energy difference observed between $E_D$ and $E_L$ in Fig. 3c, providing evidence that the Fermi level is pinned to the molecular LUMO level 140 meV below $E_D$. This interpretation is quantitatively supported by the increased width of the pinned region (in $V_G$) as molecular concentration is increased (i.e., since higher molecular concentrations can store more charge). For example, the $E_D$ plateau at a fixed molecular concentration of 4.3 x $10^{12}$ cm$^{-2}$ has a width of $\Delta V_G = 50 \pm 5$ V, which corresponds to a surface charge density of $\Delta \sigma = (3.9 \pm 0.4) \times 10^{12}$ e$^-$/cm$^2$ that reasonably matches the molecular concentration (charge is calculated using the capacitance value acquired via Eq. 1).

The Fermi level pinning described above for the *static* molecular configurations of Fig. 4a is intimately related to the *dynamic* molecular reconfigurations that enable molecular concentration to be continuously tuned by gate voltage, the central focus of this paper. When the graphene Fermi level is securely pinned by molecular LUMO states, new electrons added to the device (e.g., by an increase in $V_G$) do not cause the Fermi level to rise in energy since LUMO levels absorb any new charge added to the graphene at $E_F$. On the other hand, if there are not enough molecules on the surface to pin the Fermi level, then increasing the gate voltage causes electrons to occupy graphene band states at energies *higher* than $E_L$. This is the origin (from an energetic perspective) of the force that drives the molecules to *move* on the surface in order to dynamically change the molecular concentration when $V_{G\text{-set}}$ is modified under "diffusive conditions" (i.e., when $I_{SD} \neq 0$). When $V_{G\text{-set}}$ is increased under diffusive conditions then the molecular concentration must also increase to *maintain* Fermi level pinning (the overall lowest energy state) so as to enable charge to flow into lower-energy LUMO levels rather than higher-energy Dirac band states.

These concepts allow us to formulate a simple model for predicting the expected concentration of molecules on the graphene surface for a given gate voltage. We start with the assumption that the lowest-energy electronic configuration under diffusive conditions (and when -10 V < $V_G$) occurs when the Fermi energy is pinned at $E_L$. For a given value of $V_{G\text{-set}}$ the total charge density on the molecule-decorated FET surface ($\sigma_T = -C V_{G\text{-set}}$) will have contributions both from charge carried by the molecules ($\sigma_M$) and charge carried by the graphene Dirac band ($\sigma_G$):

$$-C V_{G\text{-set}} = \sigma_M + \sigma_G . \qquad (2)$$

If each charged molecule contains one electron in its LUMO state (assuming that double occupancy is forbidden due to the large Hubbard energy of the LUMO state[8]) then the total molecular charge is $\sigma_M = -n_M$ where $n_M$ is the surface concentration of molecules and $\sigma_M$ has units of |e|. Eq. 2 then leads to the following expression for $n_M$:

$$n_M = C V_{G\text{-set}} + \sigma_G . \qquad (3)$$

Because $E_F$ is pinned at $E_L$ by the molecular coverage, and $E_L$-$E_D$ < 0, $\sigma_G$ can be found by integrating the density of states in the graphene band from the Dirac point to $E_L$, resulting in the following well-known expression:[22]

$$\sigma_G = \frac{|E_D - E_L|^2}{\pi \hbar^2 v_F^2} \quad . \tag{4}$$

Combining Eqs. 4 and 3 leads to the final expression for molecular concentration as a function of $V_{G\text{-set}}$:

$$n_M = CV_{G-set} + \frac{|E_D - E_L|^2}{\pi \hbar^2 v_F^2} \quad . \tag{5}$$

Using $v_F$ = 1.1 x $10^6$ m/s, we are able to fit Eq. 5 to the $n_M$ vs. $V_{G\text{-set}}$ data of Fig. 2g by using |$E_D$ - $E_L$| and C as fitting parameters. Eq. 5 fits the data well for a value of |$E_D$ - $E_L$| = 142 ± 23 meV and a value of C = (7.9 ± 0.6) × $10^{10} cm^{-2} V^{-1}$. This value of capacitance agrees well with our independently determined device capacitance of C = (7.8 ± 0.2) × $10^{10} cm^{-2} V^{-1}$ (Fig. S5). A consequence of the good agreement between these capacitance values is confirmation that each F4TCNQ molecule carries a single electron of charge, since every additional electron accumulated by increasing the gate voltage corresponds to an additional molecule.

We can further check the validity of this conceptual framework by comparing the value of $E_D$ - $E_L$ obtained from our molecular concentration measurements with the value obtained independently from the STS measurements shown in Figs. 3 and 4. STS enables us to obtain $E_D$ - $E_L$ in two ways: first by extracting $E_L$ and $E_D$ directly from the dI/dV$_S$ spectra in Figs. 3a and 3b and subtracting them, and second from the $E_D$ energy plateau caused by Fermi level pinning in Fig. 4. Using the first method we see from Fig. 3 that for $V_G$ = -30V our on-molecule dI/dV$_S$ spectrum yields $E_L$ = 165 meV while our off-molecule spectrum yields $E_D$ = 305 meV. This results in a value of $E_D$ - $E_L$ = 140 ± 20 meV (the average value over the gate-voltage range -60

V < $V_G$ < -10 V is $E_D$ - $E_L$ = 143 ± 9 meV which remains quite close to this value). Using the second method, the Fermi-level pinning data of Fig. 4 reveals a Dirac point plateau at $E_D$ – $E_F$ = 140 meV ± 5 meV. Since $E_F$ is pinned at $E_L$ under these conditions, this reflects a value of $E_D$ – $E_L$ = 140 meV ± 5 meV. Both methods are in agreement with the value |$E_D$ - $E_L$| = 142 ± 23 meV obtained from the concentration-based analysis of Eq. 5, and thus support the overall physical picture that we have presented.

One consequence of this analysis is that measurement of molecular concentration on a graphene FET is shown to provide a new method for quantitatively determining the energy of molecular frontier orbitals with respect to the graphene Dirac point (i.e., $E_D$ - $E_L$). This new method is potentially valuable for determining the energy alignment of highly mobile adsorbates since it can be extremely difficult to prevent them from moving when they are under an STM tip during the bias sweeps required for STS (characterizing small devices can also be quite challenging for X-ray-based probes). In our previous work, for example, we found it necessary to anchor $F_4TCNQ$ molecules to a secondary immobile molecular template in order to probe their LUMO levels in the charged state via STS.[8] Such molecular templating, however, can alter local dielectric environments and influence molecular orbital energies.[23,24] Our new technique of measuring gate-dependent molecular concentrations allows one to bypass templating and to access molecular energy-level information via a completely different method.

In conclusion, we have demonstrated that molecular concentration at the surface of a graphene FET can be continuously and reversibly manipulated via a back-gate voltage applied simultaneously with source-drain current. The equilibrium molecular concentration is precisely determined by the capacitance between the back-gate electrode and the graphene, in combination with the energy difference between the Dirac point and the molecular LUMO level. The driving force behind this dynamic mechanical reconfiguration of molecular concentration is the energetic favorability of molecular Fermi-level pinning compared to filling graphene Dirac bands. The energy alignment of the molecular LUMO level obtained from a concentration-based analysis using these concepts compares well with the value determined from STS.

Supporting information including experimental methods, characteristics of the graphene device and spectra fitting methodology is available free of charge via the internet at http://pubs.acs.org.

**Acknowledgments**

This research was supported by the Director, Office of Science, Office of Basic Energy Sciences, Materials Sciences and Engineering Division, of the US Department of Energy under contract no. DE-AC02-05CH11231 (Nanomachine program-KC1203) (STM imaging and spectroscopy), by the Molecular Foundry (graphene growth, growth characterization), by the National Science Foundation grant DMR-1807233 (device fabrication). K.W. and T.T. acknowledge support from the Elemental Strategy Initiative conducted by the MEXT, Japan ,Grant Number JPMXP0112101001 (characterization of BN crystals) and JSPS KAKENHI Grant Number JP20H00354 (growth of BN crystals). A.R. acknowledges funding by the Deutsche Forschungsgemeinschaft (DFG, German Research Foundation) – 453903355 (data analysis). F.L. acknowledges support from Kavli ENSI Philomathia Graduate Student Fellowship.

**References:**

1. Feenstra, R. M. & Martenssson, P., "Fermi-Level Pinning at the Sb/GaAs(110) Surface Studied by Scanning Tunneling Spectroscopy", *Physical Review Letters* **61**, 447–450 (1988).
2. Spicer, W. E., Newman, N., Spindt, C. J., Liliental-Weber, Z. & Weber, E. R., """Pinning"' and Fermi level movement at GaAs surfaces and interfaces", *Journal of Vacuum Science & Technology A: Vacuum, Surfaces, and Films* **8**, 2084–2089 (1990).
3. Grassman, T. J., Bishop, S. R. & Kummel, A. C., "An atomic view of Fermi level pinning of Ge(100) by O2", *Surface Science* **602**, 2373–2381 (2008).
4. Hurdax, P., Hollerer, M., Puschnig, P., Lüftner, D., Egger, L., Ramsey, M. G. & Sterrer, M., "Controlling the Charge Transfer across Thin Dielectric Interlayers", *Advanced Materials Interfaces* **7**, 1–7 (2020).
5. Sharbati, M. T., Du, Y., Torres, J., Ardolino, N. D., Yun, M. & Xiong, F., "Low-Power, Electrochemically Tunable Graphene Synapses for Neuromorphic Computing", *Advanced Materials* **30**, 1–6 (2018).
6. Cui, K., Ivasenko, O., Mali, K. S., Wu, D., Feng, X., Müllen, K., De Feyter, S. & Mertens, S. F. L., "Potential-driven molecular tiling of a charged polycyclic aromatic compound", *Chemical Communications* **50**, 10376–10378 (2014).
7. Cai, Z. F., Yan, H. J., Wang, D. & Wan, L. J., "Potential- and concentration-dependent self-assembly structures at solid/liquid interfaces", *Nanoscale* **10**, 3438–3443 (2018).
8. Wickenburg, S., Lu, J., Lischner, J., Tsai, H.-Z., Omrani, A. A., Riss, A., Karrasch, C., Bradley, A., Jung, H. S., Khajeh, R., Wong, D., Watanabe, K., Taniguchi, T., Zettl, A., Neto, A. H. C., Louie, S. G. & Crommie, M. F., "Tuning charge and correlation effects for a single molecule on a graphene device", *Nature Communications* **7**, 13553 (2016).
9. Pinto, H., Jones, R., Goss, J. P. & Briddon, P. R., "P-type doping of graphene with F4-TCNQ", *Journal of Physics Condensed Matter* **21**, 23–26 (2009).
10. Barja, S., Garnica, M., Hinarejos, J. J., Vázquez De Parga, A. L., Martín, N. & Miranda, R., "Self-organization of electron acceptor molecules on graphene", *Chemical Communications* **46**, 8198-8200 (2010).
11. Khomyakov, P. A., Giovannetti, G., Rusu, P. C., Brocks, G., Van Den Brink, J. & Kelly, P. J., "First-principles study of the interaction and charge transfer between graphene and metals", *Physical Review B - Condensed Matter and Materials Physics* (2009). doi:10.1103/PhysRevB.79.195425
12. Kumar, A., Banerjee, K., Dvorak, M., Schulz, F., Harju, A., Rinke, P. & Liljeroth, P., "Charge-Transfer-Driven Nonplanar Adsorption of F4TCNQ Molecules on Epitaxial Graphene", *ACS Nano* **11**, 4960–4968 (2017).
13. Coletti, C., Riedl, C., Lee, D. S., Krauss, B., Patthey, L., Von Klitzing, K., Smet, J. H. & Starke, U., "Charge neutrality and band-gap tuning of epitaxial graphene on SiC by molecular doping", *Physical Review B - Condensed Matter and Materials Physics* **81**, 1–8 (2010).
14. Akiyoshi, H., Goto, H., Uesugi, E., Eguchi, R., Yoshida, Y., Saito, G. & Kubozono, Y., "Carrier Accumulation in Graphene with Electron Donor/Acceptor Molecules", *Advanced Electronic Materials* **1**, 1–5 (2015).
15. Lu, J., Tsai, H.-Z., Tatan, A. N., Wickenburg, S., Omrani, A. A., Wong, D., Riss, A., Piatti, E., Watanabe, K., Taniguchi, T., Zettl, A., Pereira, V. M. & Crommie, M. F.,

"Frustrated supercritical collapse in tunable charge arrays on graphene", *Nature Communications* **10**, 477 (2019).
16. Tsai, H.-Z., Lischner, J., Omrani, A. A., Liou, F., Aikawa, A. S., Karrasch, C., Wickenburg, S., Riss, A., Natividad, K. C., Chen, J., Choi, W.-W., Watanabe, K., Taniguchi, T., Su, C., Louie, S. G., Zettl, A., Lu, J. & Crommie, M. F., "A molecular shift register made using tunable charge patterns in one-dimensional molecular arrays on graphene", *Nature Electronics* **3**, 598-603 (2020).
17. Tsai, H. Z., Omrani, A. A., Coh, S., Oh, H., Wickenburg, S., Son, Y. W., Wong, D., Riss, A., Jung, H. S., Nguyen, G. D., Rodgers, G. F., Aikawa, A. S., Taniguchi, T., Watanabe, K., Zettl, A., Louie, S. G., Lu, J., Cohen, M. L. & Crommie, M. F., "Molecular Self-Assembly in a Poorly Screened Environment: F4TCNQ on Graphene/BN", *ACS Nano* **9**, 12168–12173 (2015).
18. Pavlíček, N., Swart, I., Niedenführ, J., Meyer, G. & Repp, J., "Symmetry dependence of vibration-assisted tunneling", *Physical Review Letters* **110**, 1–5 (2013).
19. Qiu, X. H., Nazin, G. V. & Ho, W., "Vibronic states in single molecule electron transport", *Physical Review Letters* **92**, 1–4 (2004).
20. Zhang, Y., Brar, V. W., Wang, F., Girit, C., Yayon, Y., Panlasigui, M., Zettl, A. & Crommie, M. F., "Giant phonon-induced conductance in scanning tunnelling spectroscopy of gate-tunable graphene", *Nature Physics* **4**, 627–630 (2008).
21. Decker, R., Wang, Y., Brar, V. W., Regan, W., Tsai, H.-Z., Wu, Q., Gannett, W., Zettl, A. & Crommie, M. F., "Local Electronic Properties of Graphene on a BN Substrate via Scanning Tunneling Microscopy", *Nano Letters* **11**, 2291–2295 (2011).
22. Castro Neto, A. H., Guinea, F., Peres, N. M. R., Novoselov, K. S. & Geim, A. K., "The electronic properties of graphene", *Reviews of Modern Physics* **81**, 109–162 (2009).
23. Järvinen, P., Hämäläinen, S. K., Banerjee, K., Häkkinen, P., Ijäs, M., Harju, A. & Liljeroth, P., "Molecular self-assembly on graphene on SiO2 and h-BN substrates", *Nano Letters* **13**, 3199–3204 (2013).
24. Cochrane, K. A., Schiffrin, A., Roussy, T. S., Capsoni, M. & Burke, S. A., "Pronounced polarization-induced energy level shifts at boundaries of organic semiconductor nanostructures", *Nature Communications* **6**, 8312 (2015).

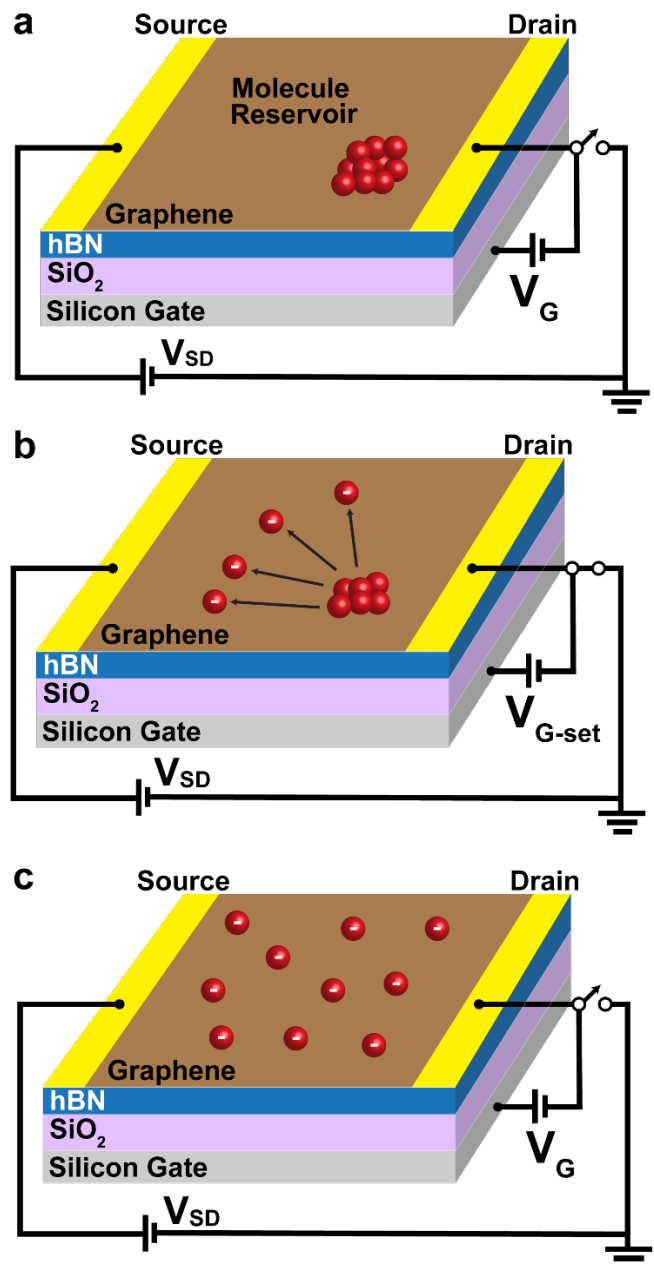

**Fig. 1 | Controlling molecular concentration on a graphene field-effect transistor (FET):** (a) Sketch representing the $F_4TCNQ$ molecular configuration (red balls) in the "as grown" state after thermal evaporation onto a graphene FET. (b) Application of a back-gate voltage, $V_{G-set}$, to the FET while simultaneously flowing a source-drain current $I_{SD}$ through the graphene causes the molecules to diffuse out onto the graphene surface. Typical values of $V_{G-set}$ applied are -60 V to 60 V. Typical $V_{SD}$ used to change the molecular concentration is 2V to 6V, and typical $I_{SD}$ is 0.5 mA to 2 mA. (c) When the source-drain current is turned off the device cools and the molecules freeze in place. The gate voltage $V_G$ is then returned to 0 V and a sample bias of $V_S$=2 V and current setpoint of 2 pA can be used for stably scanning the molecules.

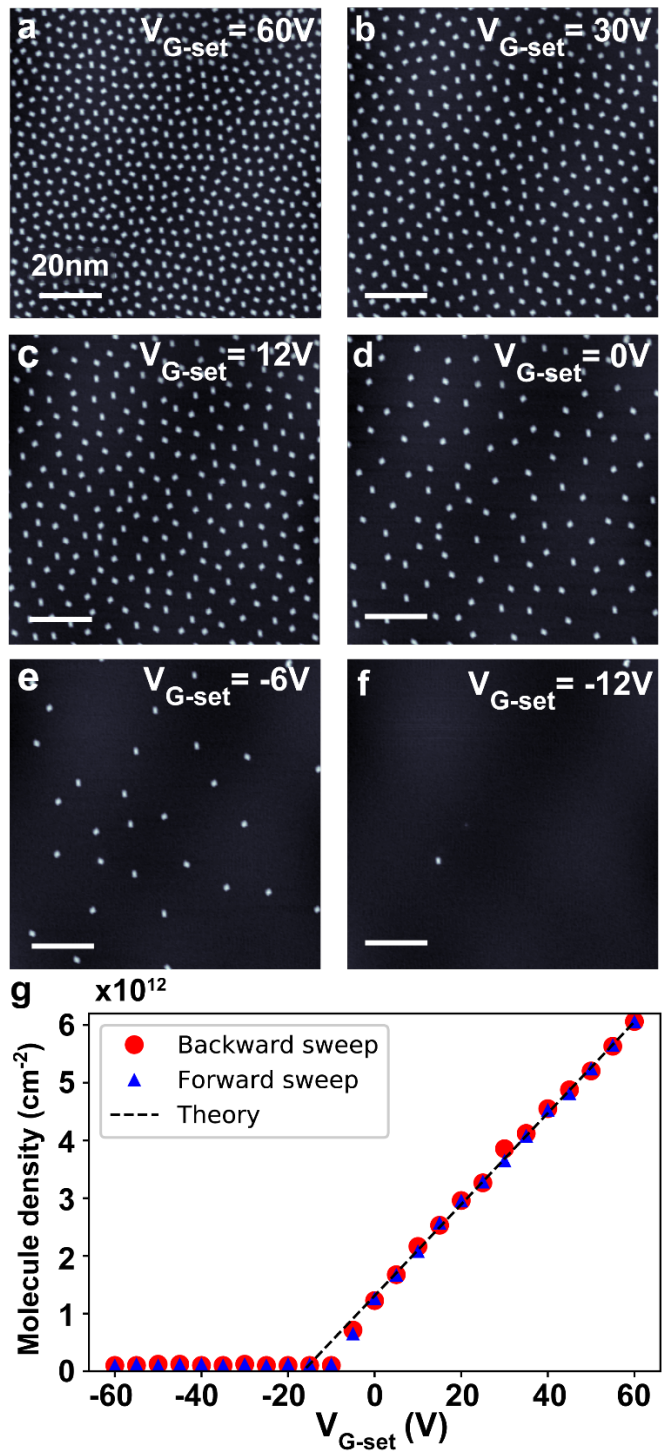

**Fig. 2 | Tuning molecular concentration by using FET gate voltage:** (a)-(f) STM topographs of the same area on the surface of a graphene field-effect transistor after tuning the molecular surface concentration with different values of gate set-voltage over the range -12 V < $V_{G-set}$ < 60 V. (g) Measured molecular concentration as a function of $V_{G-set}$. No hysteresis is observed for forward and backward sweeps. The dashed line shows a theoretical fit to the data using Eq. 5.

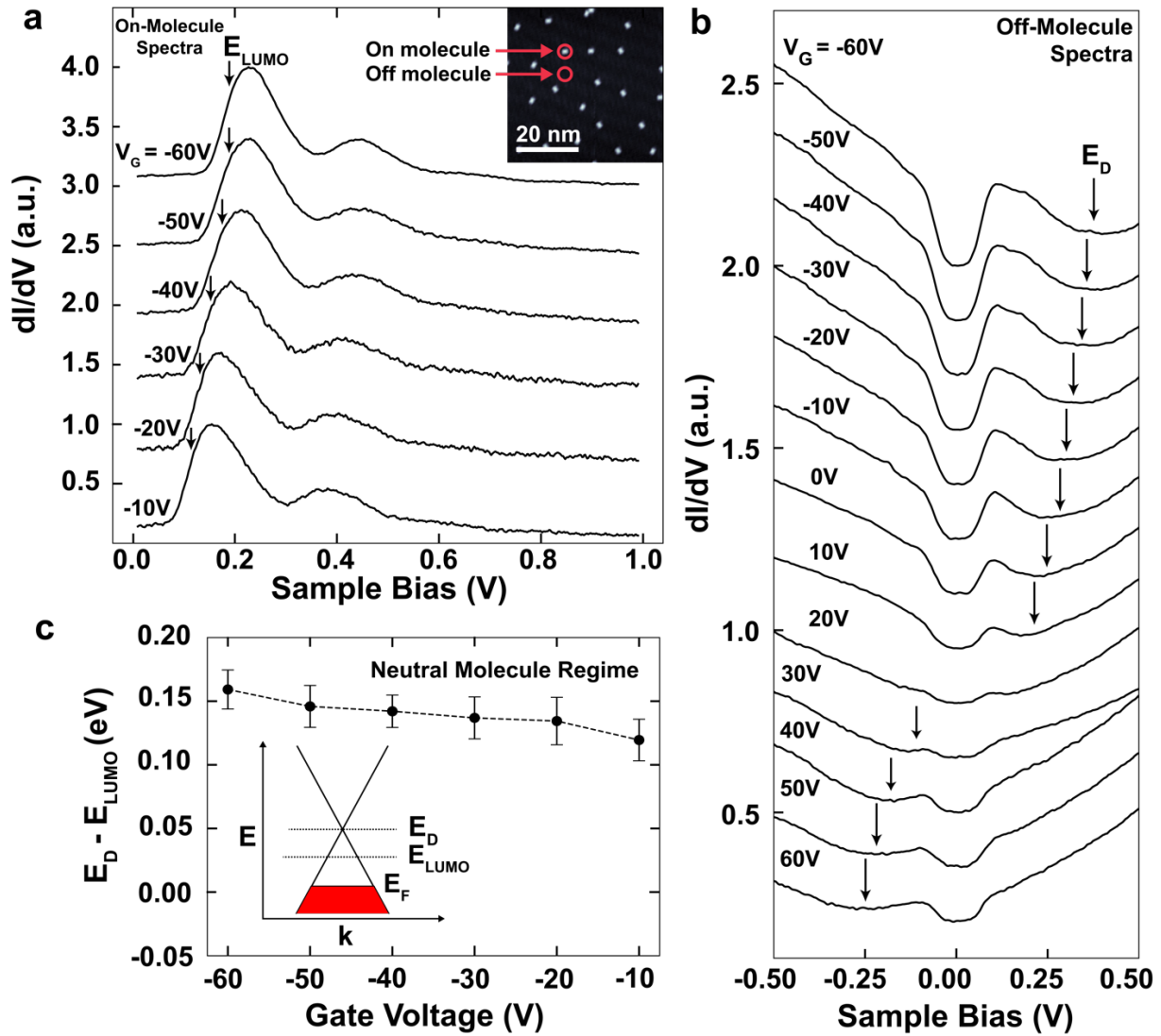

**Fig. 3 | STM spectroscopy of F$_4$TCNQ-decorated graphene field-effect transistor (FET):** (a) dI/dV$_S$ spectra measured while holding the STM tip directly above an F$_4$TCNQ molecule for a molecular concentration of 0.8 x 10$^{12}$ cm$^{-2}$ at the surface of a graphene FET (-60 V < V$_G$ < -10 V). Energy location of the lowest unoccupied molecular orbital (E$_L$) is marked. Inset shows a representative image of the surface at this molecular concentration. (b) dI/dV$_S$ spectra measured while holding the STM tip over bare patches of graphene ~100 Å away from nearby F$_4$TCNQ molecules for the same surface conditions measured in (a) (-60 V < V$_G$ < 60 V, n$_M$ = 0.8 x 10$^{12}$ cm$^{-2}$). Graphene Dirac point energy (E$_D$) is marked. Spectroscopy parameters: I$_{setpoint}$ = 50 pA, V$_{setpoint}$ = 1 V on graphene, I$_{setpoint}$ = 10 pA, V$_{setpoint}$ = 1 V on molecule. (c) Gate-voltage dependence of the Dirac point energy relative to the molecular LUMO energy (E$_D$-E$_L$) for molecular concentration n$_M$ = 0.8 x 10$^{12}$ cm$^{-2}$. Inset shows a simplified representation of the electronic structure of F$_4$TCNQ molecules on graphene for large negative gate voltages (E$_F$ = Fermi energy).

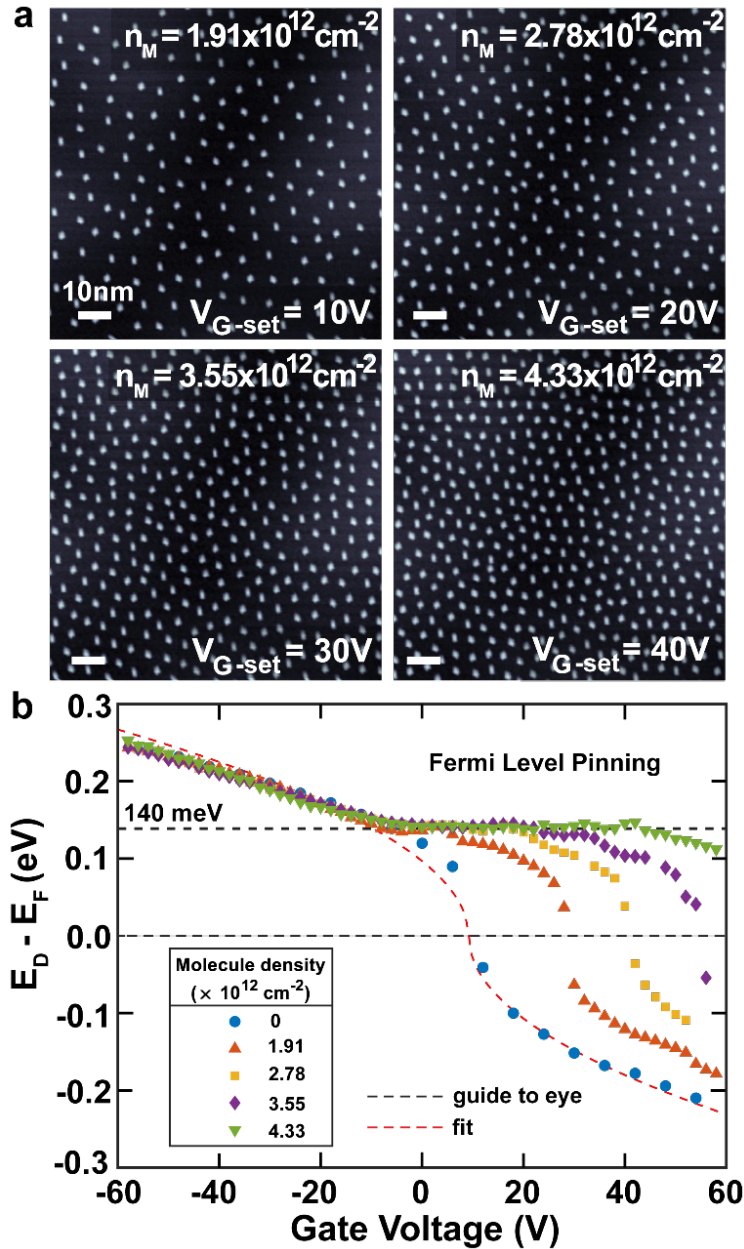

**Fig. 4 | Fermi level pinning of F$_4$TCNQ-doped graphene field-effect transistor (FET):** (a) STM images of graphene FET surface decorated with different molecular densities for measurement of molecule-induced Fermi-level pinning shown in (b). (b) Gate-voltage dependence of Dirac point energy ($E_D$) measured via STS on graphene FET surface for the different molecular densities shown in (a). A concentration-dependent plateau in $E_D$ indicates Fermi level pinning. Red dashed line shows fit of Eq. 1 to data for zero molecular concentration ($n_M = 0$).

# Supplementary Material

# Imaging reconfigurable molecular concentration on a graphene field-effect transistor

Franklin Liou[1,2,3†], Hsin-Zon Tsai[1,2†], Andrew S. Aikawa[1,2], Kyler C. Natividad[1], Eric Tang[1], Ethan Ha[1], Alexander Riss[4], Kenji Watanabe[5], Takashi Taniguchi[6], Johannes Lischner[7], Alex Zettl[1,2,3], and Michael F. Crommie[1,2,3]*

[1]Department of Physics, University of California at Berkeley, Berkeley, CA 94720, United States.

[2]Materials Sciences Division, Lawrence Berkeley National Laboratory, Berkeley, CA 94720, United States.

[3]Kavli Energy NanoSciences Institute at the University of California at Berkeley, Berkeley, CA 94720, USA.

[4]Physics Department E20, Technical University of Munich, James-Franck-Straße 1, D-85748 Garching, Germany

[5]Research Center for Functional Materials, National Institute for Materials Science, 1-1 Namiki, Tsukuba 305-0044, Japan

[6]International Center for Materials Nanoarchitectonics, National Institute for Materials Science, 1-1 Namiki, Tsukuba 305-0044, Japan

[7]Department of Materials, Imperial College London, Prince Consort Rd, London SW7 2BB, UK.

[†]F.L. and H.-Z. T. contributed equally to this paper

★Corresponding author: crommie@berkeley.edu.

**Graphene transistor fabrication**

We fabricated graphene/hBN field-effect transistors on highly doped $SiO_2$/Si by mechanical exfoliation. Electrical source and drain contacts were fabricated by depositing 3nm Cr and 10 nm Au through a stencil mask. The doped silicon substrate was used as the back-gate. The completed device geometry is illustrated in Fig. 1(a). After placement in UHV, the graphene surface was cleaned by high-temperature annealing in vacuum at 400C for 12 hours.

**Procedure to change molecular surface concentration on the graphene transistor:**

Molecular concentration was controlled by applying a set-voltage ($V_{G\text{-set}}$) to the device back-gate while a source-drain current of $I_{SD}$ ~1.5mA is *simultaneously* passed through the graphene using a typical $V_{SD}$ ~4V (resulting in a typical current density of ~40A/m) as sketched in Fig. 1. During current flow the tip is retracted by 300 nm to avoid interaction with the surface. As long as $V_{G\text{-set}}$ is above the threshold -10 V < $V_{G\text{-set}}$ (which ensures that the molecules are negatively charged[8,15,16]) then a well-defined molecular surface concentration will be established after allowing enough time for equilibration (usually ~100 s).

**STM/STS measurements**

STM/STS measurements were performed under UHV conditions at T = 4.5 K using a commercial Omicron LT STM with Pt/Ir tips. STM topography was obtained in constant-current mode. STM tips were calibrated on a Au(111) surface by measuring the Au(111) Shockley surface state before all STS measurements. STS was performed under open feedback conditions by lock-in detection of the tunnel current driven by a wiggle voltage of 6–16 mV (r.m.s.) at 401 Hz added to the tunneling bias. WSxM software was used to process all STM and AFM images.

**Protocol for finding the experimental Dirac point energy ($E_D$) from dI/dV spectra:**

In order to obtain the experimental graphene Dirac point energy ($E_D$), we perform a fit on the scanning tunneling spectroscopy (STS) dI/dV data. A full set of dI/dV spectra with different molecule concentrations is shown in Fig. S4. The Dirac point causes a dip in the dI/dV spectra, but finding its precise energy is complicated by a well-known phonon gap feature that occurs in graphene STS and which appears in all of our calibrated dI/dV spectra (this gap-like feature is caused by phonon-assisted inelastic tunneling[1]). The phonon gap feature has been shown to span the energy range -65 meV < $E$ < 65 meV and thus offsets spectral features away from the Fermi level by 65 meV.[1,2] Consequently, obtaining the true value of $E_D$ requires that this inelastic offset be taken into account. For ease of fitting, the phonon gap feature can be removed via a simple mathematical algorithm whereby we "cut out" the phonon gap and stitch together the data outside of the phonon gap region by joining the two resulting curves (i.e., the positive and negative bias branches of the dI/dV spectrum) at the Fermi level and then perform a fit on the collapsed data. Due to finite broadening of the phonon gap feature,[2] we stitch together the dI/dV branches starting at ±100 meV to eliminate the phonon gap feature entirely. After finding the energy of the Dirac-point-induced minimum in dI/dV from our fit, we then shift this energy by the known value of the phonon gap (65 meV) to obtain the correct $E_D$ value. The following protocol describes this fitting procedure in detail:

1. We first collapse the dI/dV spectrum by stitching together the dI/dV branches having

$|V_b| > 100$ meV at zero energy such that the new values of energy are assigned as $V'_b = V_b - sgn(V_b) \times 100$ meV.

2. We then find the Dirac-point-induced dip in the collapsed dI/dV spectrum about the minimum by using a Gaussian function: $y(V'_b) = -ae^{-\frac{(V'_b - b)^2}{c}} + d$.

3. We reverse the collapsing process of step 1 by adding back 100 meV to the fitted value of the parameter $b$ by using $\beta = b + sgn(b) \times 100$ meV.

4. To account for the phonon gap energy, we then subtract the known energy of 65 meV to obtain the final value of $E_D$: $E_D = \beta - sgn(b) \times 65$ meV.

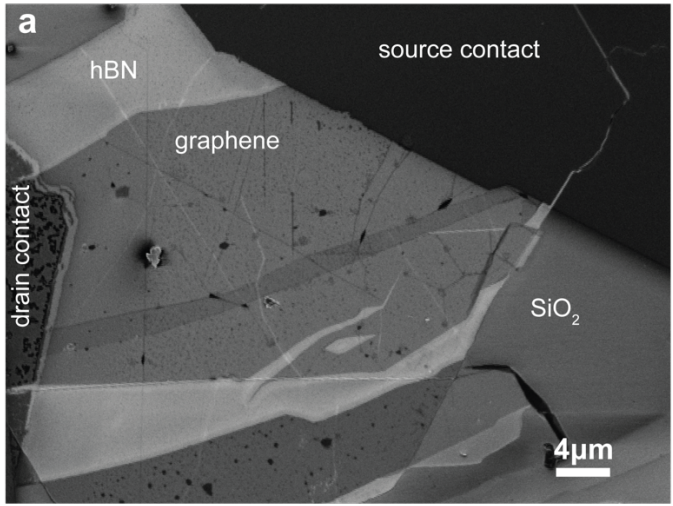

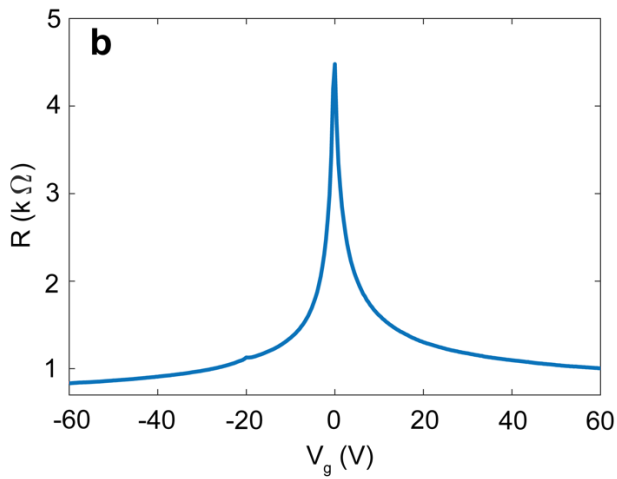

**Fig. S1 | SEM image of graphene FET and the device transconductance.** (a) An SEM image of the measured graphene device. The graphene region can be seen in the center of the hBN flake contacted by two shadow mask defined Cr/Au contacts. (b) Device transconductance as a function of applied gate voltage $V_G$. The transconductance of the pristine graphene device shows the charge neutrality point close to $V_G = 0$ V.

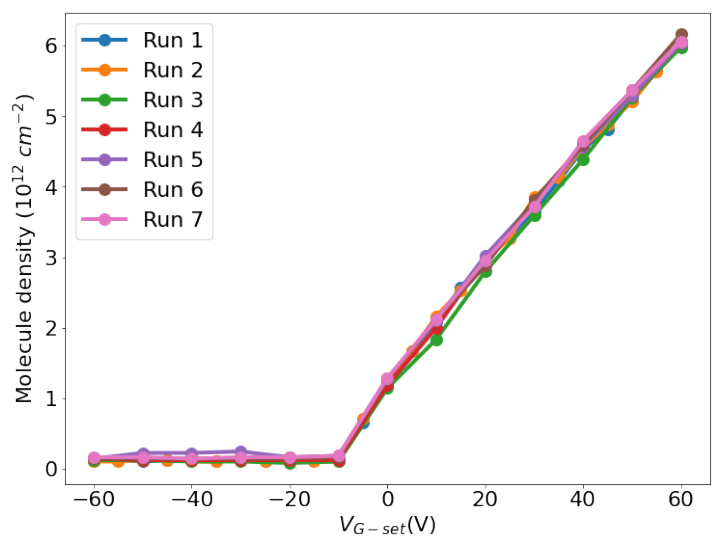

**Fig. S2 | Gate-dependent molecular surface concentration over multiple density changing operation cycles.** The gate-dependent molecular surface concentration was observed to remain consistent over many density changing operation cycles, suggesting that the molecule storage and releasing process of the molecular reservoirs is fully reversible.

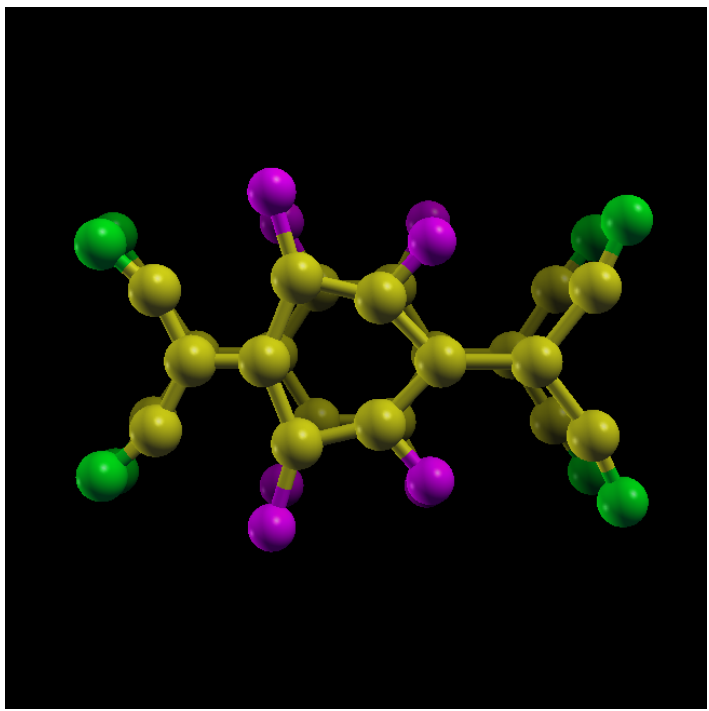

**Fig. S3 | 60 meV Phonon mode of F$_4$TCNQ.** An asymmetric phonon mode is found in F$_4$TCNQ at 60 meV from DFT calculations. This phonon offsets the STS spectrum taken on F$_4$TCNQ by 60 meV to a higher energy via an inelastic tunneling mechanism as described in ref. [1] .[3–5] The vibronic features (i.e., satellite peaks) observed in the STS spectrum taken on F$_4$TCNQ here are the same as those described in ref. [5]. The main difference between this work and the work of ref. [5] is that here we had to acquire our STM spectra with the tip held over the center of each molecule to prevent them from moving during spectroscopic measurement (since the molecules here are not anchored). In ref. [5] we were able to suppress the 60 meV inelastic effect by acquiring spectra at the extremities of each molecule since the molecules there were anchored to PCDA islands.

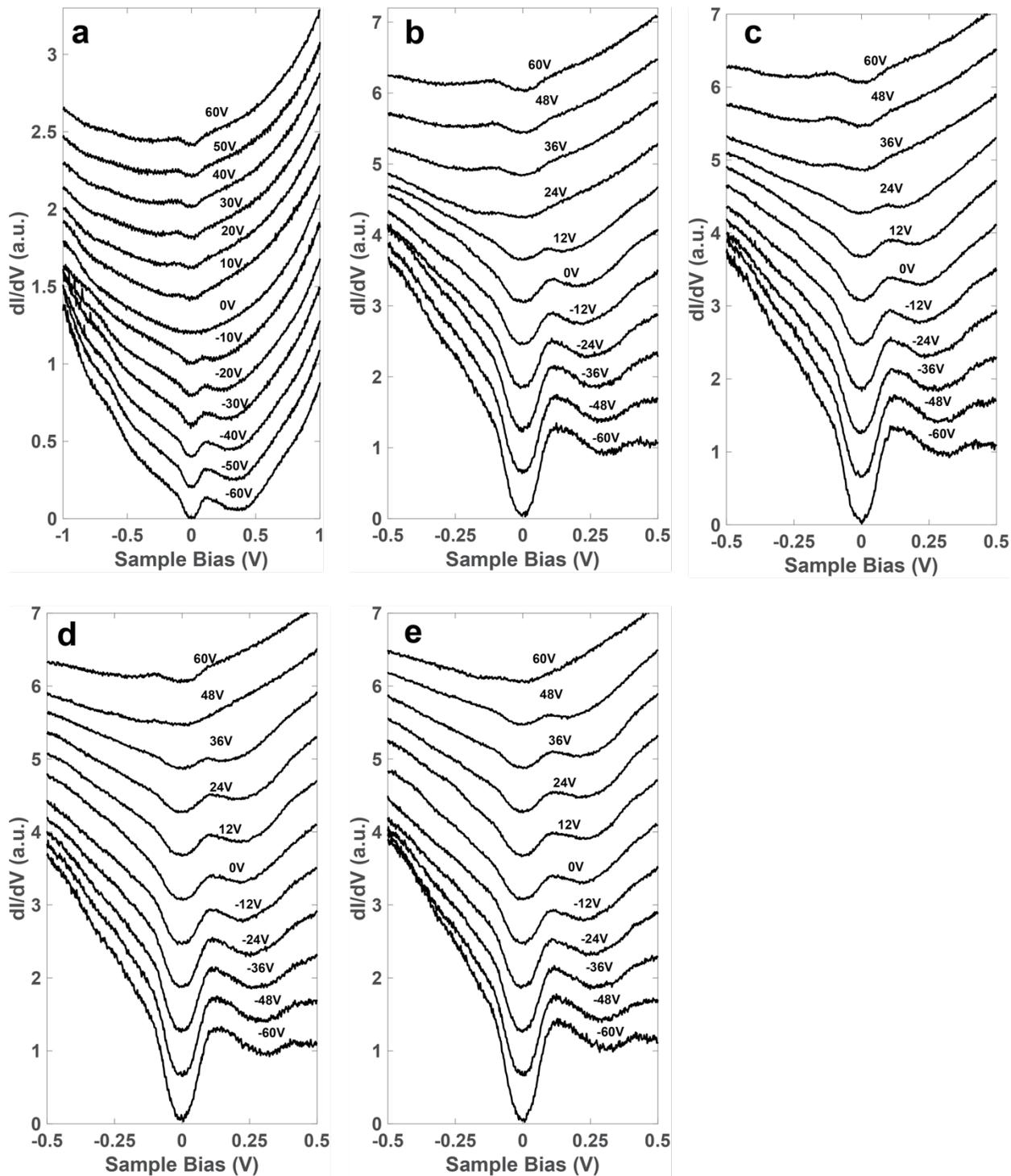

**Fig. S4 | Gate-dependent dI/dV spectra on graphene for different molecule densities.** (a) dI/dV spectra on graphene for zero molecular density. (b) dI/dV spectra on graphene with $1.91 \times 10^{12}$ molecules $/cm^2$. (c) dI/dV spectra on graphene with $2.78 \times 10^{12}$ molecules $/cm^2$. (d) dI/dV spectra on graphene with $3.55 \times 10^{12}$ molecules $/cm^2$. (e) dI/dV spectra on graphene with $4.33 \times 10^{12}$ molecules $/cm^2$.

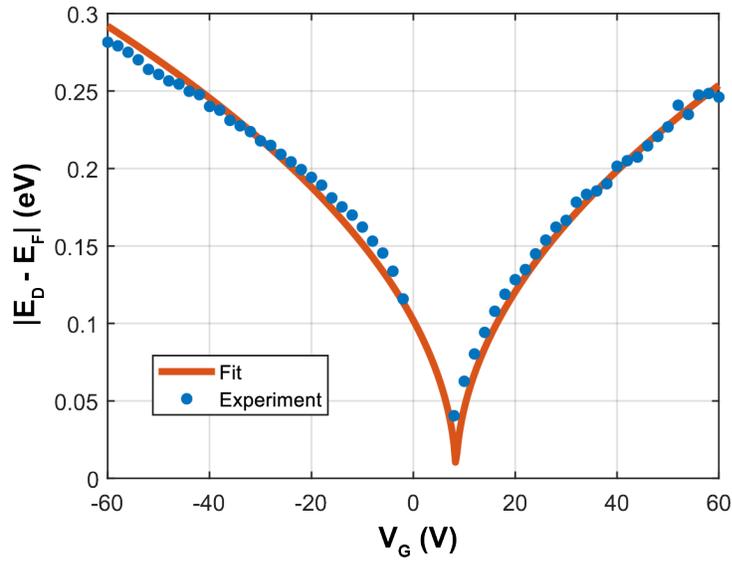

**Fig. S5 | Gate-dependence of the graphene Dirac point energy before molecule deposition.** The absolute value of the Dirac point energy relative to the Fermi level is plotted as a function of gate voltage for our pristine graphene device and shows good agreement with Eq. (1) in the main paper. A capacitance of $(7.8 \pm 0.2) \times 10^{10} cm^{-2} V^{-1}$ is extracted from the fit.

**References**
1. Y. Zhang, V. W. Brar, F. Wang, C. Girit, Y. Yayon, M. Panlasigui, A. Zettl & M. F. Crommie, "Giant phonon-induced conductance in scanning tunnelling spectroscopy of gate-tunable graphene", *Nature Physics* **4**, 627–630 (2008).
2. V. W. Brar, S. Wickenburg, M. Panlasigui, C.-H. Park, T. O. Wehling, Y. Zhang, R. Decker, Ç. Girit, A. V. Balatsky, S. G. Louie, A. Zettl & M. F. Crommie, "Observation of Carrier-Density-Dependent Many-Body Effects in Graphene via Tunneling Spectroscopy", *Physical Review Letters* **104**, 036805 (2010).
3. N. Pavliček, I. Swart, J. Niedenführ, G. Meyer & J. Repp, "Symmetry dependence of vibration-assisted tunneling", *Physical Review Letters* **110**, 1–5 (2013).
4. X. H. Qiu, G. V. Nazin & W. Ho, "Vibronic states in single molecule electron transport", *Physical Review Letters* **92**, 1–4 (2004).
5. S. Wickenburg, J. Lu, J. Lischner, H.-Z. Tsai, A. A. Omrani, A. Riss, C. Karrasch, A. Bradley, H. S. Jung, R. Khajeh, D. Wong, K. Watanabe, T. Taniguchi, A. Zettl, A. H. C. Neto, S. G. Louie & M. F. Crommie, "Tuning charge and correlation effects for a single molecule on a graphene device", *Nature Communications* **7**, 13553 (2016).